\documentclass[reprint,nobibnotes,aps,prl,amsmath,amssymb]{revtex4-2}

\usepackage{graphicx}
\usepackage{dcolumn}
\usepackage[hidelinks,breaklinks=true]{hyperref}
\usepackage{braket} 
\usepackage{subcaption}
\usepackage{romannum}
\usepackage{bm}

\begin{document}

\preprint{APS/123-QED}

\title{Bridging the Lattice Boltzmann Method and Moment Methods }

\author{Jinhua Lu}
 \email{Contact author: jinhua.LU@univ-amu.fr}
\author{Song Zhao}%
\author{Pierre Boivin}%
\author{Pierre Sagaut}
\affiliation{%
 Aix Marseille Univ, CNRS, Centrale Med, M2P2, Marseille 13013, France }%


\author{Nikolaus A. Adams}
\affiliation{%
Department of Mechanical Engineering, Chair of Aerodynamics and Fluid Mechanics, Technical University of Munich, Boltzmannstraße 15, 
Garching 85748, Germany 
}%
\affiliation{%
Munich Institute of Integrated Materials, Energy and Process Engineering, Technical University of Munich, Lichtenbergstr. 4a, Garching 85748, Germany 
}%

\date{\today}

\begin{abstract}
We derive discrete moment equations exactly equivalent to the lattice Boltzmann method (LBM) and identify the general moment closure shared by kinetic methods based on the discrete-velocity Boltzmann equation. This closure provides a route to correct lower-order dynamics without requiring accurate reconstruction of the higher-order moments, thereby substantially broadening the scope of existing closure strategies that seek increasingly accurate representations of these moments. These advances establish a unified moment framework bridging LBM and moment methods. The framework enables efficient simulations across all Knudsen regimes, overcoming the prohibitive complexity of existing moment methods at high Knudsen numbers. It further provides a common basis for integrating and extending key features of diverse kinetic methods.

\end{abstract}

\maketitle
\renewcommand{\thepage}{\arabic{page}}
The Boltzmann equation provides a fundamental kinetic description connecting molecular dynamics to macroscopic nonequilibrium phenomena:
\begin{equation}
    \partial_t f(\mathbf{x},\boldsymbol{\xi},t)
    + \xi_{\alpha}\partial_{\alpha}f(\mathbf{x},\boldsymbol{\xi},t)
    = \Omega \left( f(\mathbf{x},\boldsymbol{\xi},t) \right),
    \label{eq1}
\end{equation}
where $f$ is the particle distribution function; $\mathbf{x}$, $\boldsymbol{\xi}$, and $t$ denote physical position, particle velocity, and time, respectively; and $\Omega ()$ is the collision operator. For realistic gases, the collision operator is generally complicated and is therefore commonly approximated by the Bhatnagar-Gross-Krook (BGK) model \cite{bhatnagar1954model}
\begin{equation}
    \Omega(f)
    = -\left(f-f^{\mathrm{eq}}\right)/ \tau,
    \label{eq2}
\end{equation}
where $f^{\text{eq}}$ denotes the local Maxwellian equilibrium distribution and $\tau$ is the relaxation time. The Boltzmann equation recovers the conservation laws of mass, momentum, and energy, while retaining higher-order kinetic information beyond the Navier–Stokes description, such as rarefaction effects. Based on the Boltzmann equation, two main categories of kinetic methods, moment methods and discrete-velocity Boltzmann equation (DVBE)-based methods, have been developed for complex flows \cite{succi2018lattice,aidun2010lattice}, including dense fluids \cite{karlin2026practical}, turbulence \cite{bosch2015entropic}, and rarefied gas flows \cite{guo2013discrete}.

Moment methods solve moment equations extracted from the Boltzmann equation. They provide an effective way to explore the explicit evolution laws of high-order nonequilibrium moments. The key issue is the moment closure of the Boltzmann equation, i.e., how to determine the higher-order moments beyond selected primary moments, because the convection term of the evolution equation, Eq. \eqref{eq1}, always introduces undetermined higher-order moments.

In the successive Chapman-Enskog approximation \cite{chapman1990mathematical}, closure is obtained by expressing higher-order nonequilibrium moments as asymptotic expansions in gradients of the hydrodynamic variables. However, the resulting classical Burnett and super-Burnett equations can become ill-posed, raising questions about the validity of Chapman-Enskog solutions to the Boltzmann equation beyond the Navier-Stokes level \cite{ernst1978kinetic}. Mode-coupling theories \cite{ernst1975nonanalytic,ernst1978kinetic} and numerical simulations \cite{standish1999numerical} indicate that certain transport coefficients associated with the Burnett and super-Burnett orders diverge in the thermodynamic limit, implying nonanalytic gradient corrections that cannot be represented by a regular Chapman–Enskog expansion.

Alternatively, Grad \cite{grad1949kinetic,grad1952profile} proposed to approximate the distribution function by projecting it onto a truncated basis of Hermite polynomials to preserve the maximum exact equilibrium moments. Higher-order moments are then approximated by those of the truncated Hermite polynomials. The resulting 13-moment equations are not globally hyperbolic and may yield nonphysical solutions \cite{muller2013rational,struchtrup2002heat}. Regularized moment equations incorporating additional terms from the Chapman-Enskog expansion have been proposed to improve this issue \cite{struchtrup2003regularization}. In addition, entropy-based closures \cite{levermore1996moment} have been proposed to implicitly determine higher-order moments by minimizing the Boltzmann \(H\)-function subject to the retained moments.

The central challenges of existing moment closures are their prohibitive complexity at high Knudsen numbers (Kn), where hundreds of coupled moment equations may be required, and their insufficiently validated physical reliability in this regime. Practical applications are therefore largely limited to low-order models, such as the 13- and 26-moment systems, at moderate Kn.

Another category of kinetic methods is DVBE-based methods that restrict continuous particle velocities to a finite set and solve DVBEs by different schemes. This category includes discrete velocity model (DVM) \cite{aristov2001direct}, unified gas kinetic schemes (UGKS) \cite{xu2010unified}, discrete UGKS (DUGKS) \cite{guo2013discrete,guo2015discrete}, lattice Boltzmann method (LBM) \cite{succi2001lattice,kruger2017lattice}, and so on. Their successful applications to many high-Kn flows \cite{guo2013discrete,shi2021accuracy,xu2010unified} suggest the existence of a general and efficient moment closure strategy underlying them. An early investigation \cite{ansumali2007hydrodynamics} derived analytical solutions for rarefied planar Couette flow from continuous moment equations with moment closure determined by LBM distribution functions. The results demonstrated the feasibility of this closure for the two-dimensional nine- and sixteen-velocity Gauss–Hermite sets, denoted D2Q9 and D2Q16, respectively, in the transition regime. The analysis did not identify a general closure principle shared by DVBE-based methods or reveal how this closure operates in the fully discrete moment equations of LBM. Consequently, these results were not developed into a general moment formulation for unsteady flows, complex geometries, and high-Kn regimes.

Fully translating the success of DVBE-based methods into advances in moment methods calls for a unified moment framework that combines the general moment closure embedded in DVBE representations with explicit moment equations that preserve the underlying kinetic dynamics. This framework could substantially extend the applicability of moment methods to high-Kn flows while integrating key features of both DVBE-based kinetic methods and conventional moment methods, but remains to be established.

This Letter establishes such a framework using LBM as a representative example because of its concise formulation. We derive discrete moment equations exactly equivalent to LBM, explicitly retaining all higher-order moments involved while preserving the underlying kinetic dynamics, and identify the general closure strategy shared by all DVBE-based kinetic methods.

The construction of LBM from DVBE is briefly introduced here. Upon restricting the particle velocity space $\mathbf{\xi}$ to a discrete set $\{\mathbf{e}_i\}$ \cite{shan1998discretization}, DVBE reads
\begin{equation}
    \partial_t f_i + e_{i \alpha} \partial_\alpha f_i= \Omega_{i} \left( \ket{f} \right), \;  \text{with} \; \ket{f}_i=f_i
    \label{eq3}
\end{equation}
where the subscript $i$ represents the variables of $i$th discrete velocity.  Using an integration along the characteristic line with the trapezoidal rule, which corresponds to second-order temporal and spatial accuracy, the discrete DVBE can be expressed as
\begin{equation}
\begin{gathered}
     f_i \left( \mathbf{x} + \mathbf{e}_i \delta t, t+ \delta t\right) - f_i \left( \mathbf{x},t \right)=\\ \delta t/2 \left[ \Omega_i\left(\mathbf{x} + \mathbf{e}_i \delta t, t+ \delta t \right) 
    + \Omega_i\left(\mathbf{x} , t\right) \right] ,
\end{gathered}
    \label{eq4}
\end{equation}
where $\delta t$ is the time step.
A new distribution function $\bar{f_i}=f_i - \Omega_i \delta t/2$ is introduced to remove the explicit collision term, and Eq. \eqref{eq4} can be rewritten as
\begin{equation}
    \bar{f_i} \left( \mathbf{x} + \mathbf{e}_i \delta t, t+ \delta t\right) - \bar{f_i} \left( \mathbf{x},t \right)=\Omega_i \left( \mathbf{x},t \right) .
    \label{eq5}
\end{equation}
LBM is a Lagrangian framework to solve Eq. \eqref{eq5}, which is generally solved through three steps:
\begin{equation}
  \text{Collision:} \quad  \bar{f_i}^\text{col}=\bar{f}_i + \Omega_i \left(\ket{\bar{f}} \right),
  \label{eq6}
\end{equation}
where the subscript "col" denotes the variable after collision. Beyond the BGK collision model in Eq. \eqref{eq2}, many advanced collision models, such as multi-relaxation-time \cite{d2002multiple}, recursive regularized \cite{coreixas2017recursive}, and entropic collision models \cite{hosseini2023entropic}, have been developed to improve the stability and accuracy of LBM.
\begin{equation}
  \text{Streaming:} \quad \bar{f_i}  \left(\mathbf{x}, t+ \delta t \right) = \bar{f_i}^\text{col} \left(\mathbf{x} - \mathbf{e}_i \delta t \right).
  \label{eq7}
\end{equation}
\begin{equation}
  \text{New variables:} \; \rho=\sum_i \bar{f}_i, \; p=\rho RT, \;
    \rho u_\alpha=\sum_i \bar{f}_i e_{i \alpha},
    \label{eq8}
\end{equation}
where $\rho$, $p$, $R$, $T$, and $u_\alpha$ are density, pressure, gas constant, temperature, and velocity, respectively. The omitted spatial and temporal arguments are $\left( \mathbf{x,t} \right)$ by default. The current discussion is based on isothermal flow, and the energy or temperature  evolution is not discussed here. 

In standard LBM, the discrete velocities are usually chosen to be on-lattice so that the nonlocal streaming step remains straightforward and conservative. The on-lattice constraint complicates the use of nonuniform meshes \cite{he1996some, he1997lattice, karlin1999lattice,bouzidi2001lattice}, which are common in practical applications, and limits the applicability of LBM to wall-bounded flows at high Kn, where appropriate off-lattice discrete sets \cite{ambrucs2016lattice} are needed to effectively capture wall-induced nonequilibrium effects. As shown below, these limitations are removed within the unified moment framework proposed in this Letter.

An example of the derivation of discrete moment equations is presented using a D2Q9 velocity set \cite{qian1992lattice} with $i=0, \cdots, 8$. By defining the moment
\begin{equation}
\begin{gathered}
        M_{x^m y^n}=\sum_i e_{i x}^m e_{i y}^n  \bar{f}_i=\braket{A^{\left( m,n\right)} | \bar{f} }, \; 
        \text{with} \\ \bra{A^{(m,n)}}_i= e_{i x}^m e_{i y}^n, \; \ket{\bar{f}}_i=\bar{f}_i, \; e_{ix},e_{iy}\in[-1,0,1] c,
\end{gathered}
    \label{eq9}
\end{equation}
 where $c$ is the lattice speed related to the grid spacing $\delta x$ by $c=\delta x/ \delta t$, the 9 distribution functions can be projected onto a set of 9 primary moments
    \begin{equation}
    \begin{gathered}
    \ket{M} = [P] \ket {\bar{f}}, \quad \text{with} \quad \ket{M}_j =M_{x^m y^n}, \quad
    [P]_{ji} = e_{i x}^m e_{i y}^n, \\
    j=mN+n, \quad m,n =0, 1, \dots, N-1,
    \end{gathered}
    \label{eq10}
    \end{equation}
where $N=3$,  $\ket{M }$ is the vector of primary moments, $[P]$ is the projection matrix. Since the projection matrix is invertible, the 9 DVBEs can be recast as 9 primary moments, without loss of information.

The collision step of LBM in the moment space is
\begin{equation}
       \ket{M^\text{col}}  = \ket{M} + \Omega \left( \ket{M} \right).
       \label{eq11}
\end{equation}
Applying a Taylor expansion to $\bar{f}_i^\text{col} \left(\mathbf{x} - \mathbf{e}_i \delta t \right)$ 
\begin{equation}
\begin{gathered}
    \bar{f}_i^\text{col} \left( \mathbf{x} - \mathbf{e}_i \delta t\right)  =  \sum_{\substack{p,q=0}}^{\infty} \frac{ \left(-e_{i x} \delta t\right)^p \left(-e_{i y} \delta t\right)^q } {p!q!} \partial_x^p \partial_y^q  \bar{f}_i^\text{col},  
\end{gathered}
  \label{eq12}
\end{equation}
and then computing the primary moments, the continuous moment equations of the streaming process are 
\begin{equation}
\begin{gathered}
   M_{x^m y^n} \left(\mathbf{x}, t+ \delta t\right) = \sum_{\substack{p,q=0}}^{+ \infty} \frac{ \left(- \delta t\right)^{\left( p+q \right)} } {p!q!}
  \partial_x^p \partial_y^q  M_{x^{\left(m+p\right)} y^{\left(n+q\right)}}^\text{col}.
\end{gathered}
\label{eq13}
\end{equation}
Eqs. \eqref{eq11} and \eqref{eq13} are equivalent moment equations of LBM in spatially continuous form.

It should be noted that LBM is an intrinsically discrete algorithm, whose key properties are closely tied to its specific discretization scheme. Therefore, its equivalent moment equations in discrete form are derived as well. For an arbitrary scalar field $\phi$ sampled on the $3\times 3$ uniform stencil $\{\mathbf{x}-\mathbf{e}_i\delta t\}$, the stencil values can determine a set of approximated derivatives $\ket{\partial}$ at $\mathbf{x}$ by
\begin{equation}
\begin{gathered}
        \ket{\phi}=[W]\ket{\partial}, \; \text{with} \\
     \ket{\phi}_i=\phi(\mathbf{x} -\mathbf{e}_i\delta t), \quad [W]_{ij}= \frac{ \left(- e_{ix} \delta t \right)^p \left( -e_{iy} \delta t \right)^q}{p!q!}, \\
    \ket{\partial}_j=\underline{\left( \partial_x^p \partial_y^q \phi \right)}, \quad j=pN+q, \quad p, q \in [0,\cdots,N-1].
\end{gathered}
    \label{eq14}
\end{equation}
Here, $\underline{\left( \partial_x^p \partial_y^q \phi \right)}$ denotes the approximated derivatives that can be uniquely determined from the specified stencil. Since the coefficient matrix $[W]$ is nonsingular, this transformation is invertible and preserves all information carried by the stencil. Setting $f^\text{col}_i$ as $\phi$ in Eq. \eqref{eq14}, we have
\begin{equation}
\begin{gathered}
    \bar{f}_i^\text{col} \left( \mathbf{x} - \mathbf{e}_i \delta t\right)  =  \sum_{\substack{p,q=0}}^{N-1} \frac{ \left(-e_{i x} \delta t\right)^p \left(-e_{i y} \delta t\right)^q } {p!q!} \underline{\left(\partial_x^p \partial_y^q  \bar{f}_i^\text{col} \right)} .  
\end{gathered}
  \label{eq15}
\end{equation}
The continuous higher-order exact derivatives in Eq. \eqref{eq12} are involved in the approximated lower-order derivatives, for instance,
\begin{equation}
\begin{gathered}
     \underline{\left( \partial_x  \bar{f}_i^\text{col} \right)} = \frac{ \bar{f}_i^\text{col} \left(\delta x, 0 \right) -  \bar{f}_i^\text{col}
     \left(-  \delta x, 0 \right)}{2 \delta x} \\
    =\partial_x  \bar{f}_i^\text{col} + \sum_{n=1}^{+ \infty} \frac{ \delta x^{2n} }{\left(2n+1 \right)!} \partial_x^{2n+1}  \bar{f}_i^\text{col} ,
\end{gathered}
\label{eq16}
\end{equation}
where the local position $\mathbf{x}$ is set as $\left(0,0 \right) $. Substituting Eq. \eqref{eq15} into Eq. \eqref{eq7} and computing the primary moments yields the discrete moment equations of the streaming process
\begin{equation}
\begin{gathered}
   M_{x^m y^n} \left(\mathbf{x}, t+ \delta t\right) =M^\text{col}_{x^m y^n} + \\ \sum_{\substack{p,q=0\\(p,q)\neq(0,0)}}^{N-1} \frac{ \left(- \delta t\right)^{\left( p+q \right)} } {p!q!}
\underline{ \left( \partial_x^p \partial_y^q  M_{x^{\left(m+p\right)} y^{\left(n+q\right)}}^\text{col}\right) }.
\end{gathered}
\label{eq17}
\end{equation}
Eqs. \eqref{eq11} and \eqref{eq17} are discrete moment equations exactly equivalent to LBM at the fully discrete level. The explicit presence of all higher-order moments is the precondition for revealing its underlying moment closure and distinguishes these equations from other moment equations derived from LBM \cite{dubois2008equivalent,bellotti2022finite,fuvcik2021equivalent}, which generally modify the implicit moment closure. The derivation also applies to higher-order and three-dimensional LBM models.

The discrete moment equations have two characteristics that are essential for stable simulations across all Kn regimes. First, it corresponds to a Lax-Wendroff-type scheme that transforms second-order temporal derivatives to second-order spatial derivatives, yielding physically consistent dissipation terms for stabilizing computations. For example, the zeroth-order moment equation of LBM is
\begin{equation}
\begin{aligned}
    \rho \left( t + \delta t\right)= &\rho - \delta t \partial_\alpha\left( \rho u_\alpha\right)  + \frac{\delta t^2}{2}  \partial_\alpha \partial_\beta \left( p\delta_{\alpha \beta} +\rho u_\alpha u_\beta\right) \\ + O \left( \partial^3\right) 
    = & \rho - \delta t \partial_t \rho  + \frac{\delta t^2}{2}  \partial_t^2 \rho + O \left( \delta t^3\right).
\end{aligned}
\label{eq18}
\end{equation}
 Second, the temporal integration in Eq. \eqref{eq4} considers particle convection and collision effects simultaneously, which is essential for capturing continuum flows \cite{xu2010unified}, where the collision time scale is much shorter than the flow time scale.

Based on the derived moment equations of LBM, the moment closure underlying LBM is provided here. According to the definition of moments, Eq. \eqref{eq9}, the involved higher-order moments beyond primary ones can be determined by the distribution functions, as well as the corresponding primary moments
    \begin{equation}
    \begin{gathered}
    M_{x^m y^n}  = \braket{A^{(m,n)} | \bar{f}_i } = \bra{A^{(m,n)}} [P]^{-1} \ket{M}, \\ 
    \text{for} \quad m > N-1 \quad \text{or} \quad  n > N-1.
    \end{gathered}
    \label{eq19}
    \end{equation}
As $\bra{A^{(m,n)}} [P]^{-1}$ is a constant vector for specified $m$ and $n$, the higher-order moments $ M_{x^m y^n}$ are linear combinations of primary moments. Three remarks on the identified moment closure of LBM are in order:

Remark 1. The moment closure rests on two ingredients: the definition of moments, Eq. \eqref{eq9}, and the invertible transformation between the distribution functions and a set of linearly independent primary moments, Eq. \eqref{eq10}. Both are valid for a finite discrete velocity set. Consequently, the closure strategy underlying LBM also applies to DVBE, i.e., Eq. \eqref{eq3}, and other DVBE-based kinetic methods. The closed moment equations of DVBE for two-dimensional velocity sets can be written as
\begin{equation}
     \partial_t M_{x^m y^n}
     + \partial_x M_{x^{m+1} y^n}
     + \partial_y M_{x^m y^{n+1}}
     = \Omega \left( \ket{M} \right),
    \label{eq20} 
\end{equation}
where $M_{x^m y^n}=\braket{A^{\left( m,n\right)} | f}$, higher-order moments $ M_{x^{m+1} y^n}$ and $M_{x^m y^{n+1}}$ are determined by the general moment closure, Eq. \eqref{eq19}. The ranges of $m$ and $n$ are determined by the linearly independent primary moment set associated with the chosen discrete velocity set, and $j$ denotes the corresponding moment index.

Remark 2. Different discrete velocity sets correspond to different moment closures. Examples include velocity sets based on half-range Gauss-Hermite quadrature nodes \cite{ambrucs2016lattice}, which accurately capture wall effects; full-range Gauss-Hermite quadrature nodes \cite{kallikounis2022particles}, which achieve high-order integration accuracy with a minimum number of discrete velocities; and Newton-Cotes quadrature nodes \cite{guo2015discrete}, which are suitable for highly nonequilibrium effects. Their successful applications in various Kn regimes \cite{shi2021accuracy,guo2013discrete,guo2015discrete} demonstrate the generality of the identified moment closure for flexible discrete velocity sets. The general principle for choosing discrete velocity sets is to provide sufficient resolution in the high-probability regions of particle velocity space. The closures used in LBM with Gauss-Hermite velocity sets \cite{ansumali2007hydrodynamics} can therefore be viewed as special cases of the present moment-closure perspective.

Remark 3. The higher-order moments expressed as linear combinations of the primary moments in Eq. \eqref{eq19} generally deviate from their exact values. Discrete velocity sets based on Gauss-Hermite quadratures can exactly recover certain higher-order equilibrium moments, but not higher-order nonequilibrium moments. Other velocity sets, such as those based on half-range Gauss-Hermite quadratures, may yield fully inexact higher-order moments. Nevertheless, successful applications of high-order LBM models and other DVBE-based methods indicate that the influence of these deviations is typically constrained along the moment hierarchy and does not dominate the physical solution. A rigorous analysis of their effects remains open. The general moment closure identified in this Letter provides a new route that does not require accurate higher-order moments but instead controls the influence of their inaccuracies on lower-order moments. Its simplicity and applicability over a wide range of Kn can be demonstrated by existing DVBE-based methods \cite{shi2021accuracy,guo2013discrete,guo2015discrete,xu2010unified} and the numerical test below. This perspective substantially broadens existing closure strategies that seek increasingly accurate higher-order moments through Chapman-Enskog expansions or physical approximations.

The discrete moment equations of LBM and their closure define a unified moment framework that bridges LBM and conventional moment methods. This framework involves the physically consistent discretization of LBM, the moment closure shared by all DVBE-based methods, and the formulation of moment equations. It can therefore incorporate key features of diverse kinetic methods and substantially advance kinetic methods. Within this framework, we develop a moment-equation-based gas-kinetic scheme (MEGKS) to demonstrate its advantages.

MEGKS is a finite-volume scheme for solving the discrete moment equations of LBM and retains the collision and streaming steps of LBM. Its local collision step is identical to that of LBM, i.e., Eq. \eqref{eq11}. The moment equations of the streaming step are rewritten in finite-volume form
\begin{equation}
\begin{gathered}
   M_{x^m y^n} \left(\mathbf{x}, t+ \delta t\right) =M^\text{col}_{x^m y^n} - \frac{\delta t}{\delta V}\sum\\  \left[ 
   \begin{aligned}
       \sum_{\substack{p=1\\q=0}}^{N-1} \frac{ \left(- \delta t\right)^{\left( p+q-1 \right)} } {p!q!}
\underline{ \left( \partial_x^{p-1} \partial_y^q  M_{x^{\left(m+p\right) }y^{\left(n+q\right)}}^\text{col} \right)}\\
 \sum_{\substack{p=0\\q=1}}^{N-1} \frac{ \left(- \delta t\right)^{\left( p+q-1 \right)} } {p!q!}
\underline{ \left(\partial_x^p \partial_{y}^{q-1}  M_{x^{\left(m+p\right)} y^{\left(n+q\right)}}^\text{col} \right) }
   \end{aligned}   
   \right]^\text{T} \cdot \left(
   \begin{aligned}
         n_x \\
         n_y  
   \end{aligned}
   \right)  \delta S, \\ 
   m,n=0,1,\dots, N-1,
\end{gathered}
\label{eq21}
\end{equation}
where $\delta V$, $\left(n_x, n_y \right)$, and $\delta S$ denote the control volume, the outward unit normal vector, and the cell-face area, respectively. Eqs.~\eqref{eq11} and \eqref{eq21} define an MEGKS for a system of $N^2$ primary moments associated with a general $N\times N$ discrete velocity set. Implementation details and the simplified high-order MEGKS are provided in
Secs.~\Romannum{2} and ~\Romannum{3} of the Supplemental Material, respectively.

To demonstrate the contribution of the unified moment framework to existing moment methods, planar Couette flow across all Kn regimes is considered. Two parallel plates of distance $H$ move horizontally at velocities $u_w$ and $-u_w$, respectively. Diffuse-scattering boundary conditions \cite{guo2013discrete} are applied on both walls, with periodic conditions in the stream direction. The modified Knudsen number $k=(\sqrt{\pi}/2)\mathrm{Kn}$ ranges from the continuum regime ($k=0.001$) to the free-molecular regime ($k=100$). $\mathrm{Kn}$ is defined as
\begin{equation}
    \text{Kn}=\frac{ \lambda_F}{H}=\frac{\mu}{p}\sqrt{\frac{\pi RT}{2}},
    \label{eq23}
\end{equation}
where $\lambda_F$ and $\mu$ are the mean free path and dynamic viscosity. respectively.

\begin{figure}[hbtp]
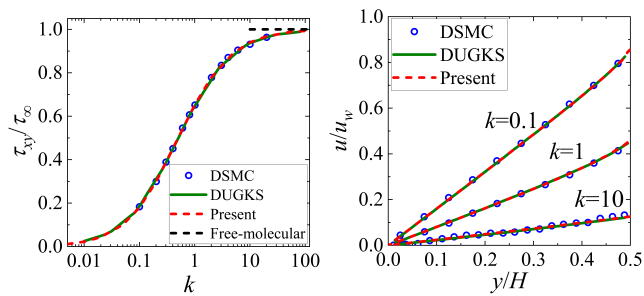

\includegraphics[width=0.48\linewidth]{Couette_stress.png}
\includegraphics[width=0.48\linewidth]{Couette_velocity.png}
\caption{\label{fig: Couette} Left: normalized shear stress $\tau_{xy}/\tau_\infty$ over full Kn regime, with $\tau_\infty=\rho u_w\sqrt{2RT/\pi}$. Right: velocity profiles at $k=0.1$, 1, and 10. Reference data are from DSMC~\cite{bahukudumbi2003unified} and DUGKS ~\cite{guo2013discrete}.}
\end{figure}

The case employs a $D2Q18^2$ discrete velocity set constructed from tensor products of half-range Gauss-Hermite quadrature nodes, leading to 324 closed moment equations. As shown in Fig. \ref{fig: Couette}, the MEGKS results agree well with DUGKS and DSMC data across all Kn regimes. Explicitly deriving and solving hundreds of coupled equations would be prohibitively complex in conventional moment methods. The present framework efficiently handles such high-order systems, providing a systematic route to moment-based simulations across all Kn regimes.

Beyond these fundamental advances in moment methods, the general moment closure and efficient simulations across all Knudsen regimes, the unified framework provides a common basis for integrating and extending key features of diverse kinetic methods.

First, when a uniform discrete-velocity set is employed, the framework is exactly equivalent to LBM and thus retains its characteristic properties on uniform meshes, including the extremely low acoustic dissipation of low-order LBM models \cite{bres2009properties}. Sec.~\Romannum{4} of the Supplemental Material provides a validation case. 

Second, by inheriting the semi-explicit LBM discretization, the unified moment framework includes an independent collision step that enables moment-dependent transport coefficients prescribed by the selected molecular collision model \cite{yan2024lattice}, which is essential for high-Mach-number rarefied gas flows. Major DVBE-based methods \cite{xu2010unified,guo2013discrete} couple particle transport and collision in a unified evolution process and commonly employ the BGK collision model to avoid significant complexity, in which all nonconserved moments relax with a single time scale. However, realistic molecular collision models generally have different relaxation rates for different moments, corresponding to distinct transport coefficients. Although extensions of the BGK model, such as the Shakhov model \cite{shakhov1968generalization} and the ellipsoidal statistical model \cite{holway1966new}, correct the Prandtl number by modifying the relaxation of the heat flux and stress, respectively, they do not permit independent relaxation rates for higher-order moments.

Third, the framework accommodates general nonuniform meshes and off-lattice velocity sets tailored to rarefied flows, such as the half-range Gauss-Hermite quadrature employed in the Couette-flow test, thereby removing the corresponding mesh and velocity-set constraints of LBM. Examples are provided in
Secs.~\Romannum{2} and ~\Romannum{3} of the Supplemental Material.

Fourth, a promising application of the unified moment framework is the use of locally adaptive velocity sets. To maximize the number of higher-order equilibrium moments recovered exactly, thereby enhancing physical fidelity while reducing the number of discrete velocities and computational cost, the Gauss-Hermite velocity set is shifted by the local flow velocity and scaled by the local temperature:
\begin{equation}
\mathbf{v}_i=\mathbf{u}_{\rm ref}
+\mathbf{e}_i\sqrt{T_{\rm ref}/T_L},
\label{eq22}
\end{equation}
where $\mathbf{e}_i$ and $\mathbf{v}_i$ denote the reference quadrature nodes and local discrete velocities, respectively; $\mathbf{u}_{\rm ref}$ and $T_{\rm ref}$ are local macroscopic velocity and temperature, respectively; and $T_L$ is the lattice temperature. Such adaptive velocity sets are adopted in the particles-on-demand LBM \cite{dorschner2018particles}, which requires repeated projections of the distribution functions between different local velocity sets under moment-invariance constraints. Similar projection issues arise in multiscale simulations \cite{kallikounis2024particles} in which velocity sets are adapted to the local Kn for improved efficiency.
\begin{figure}[hbtp]
\includegraphics[width=0.5\linewidth]{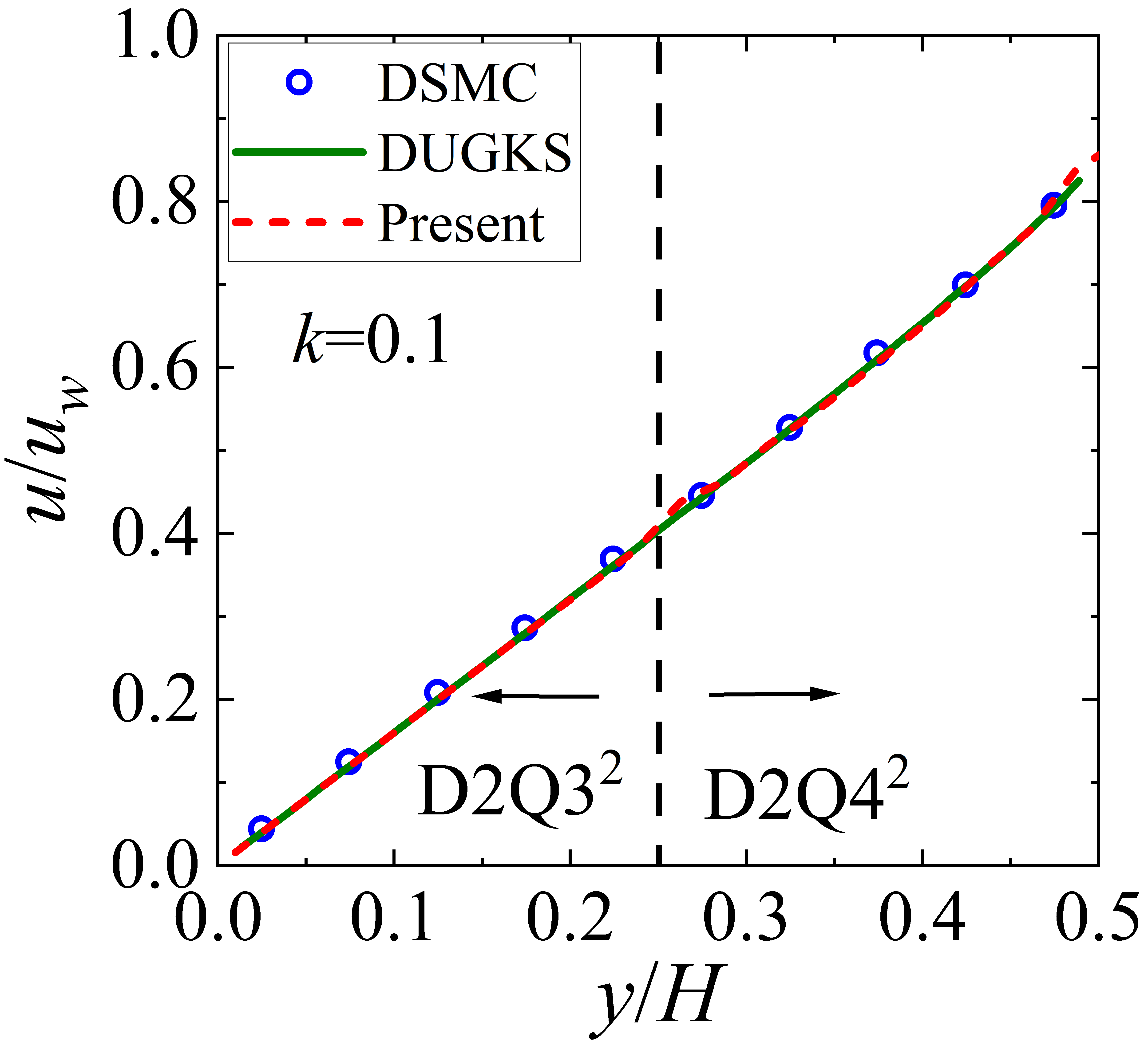}
\caption{\label{fig: Multiscale} Velocity profile from a multiscale simulation at $k=0.1$. Reference data are from DSMC~\cite{bahukudumbi2003unified} and DUGKS ~\cite{guo2013discrete}.}
\end{figure}

In the unified moment framework, information exchange between different representations can be performed directly in moment space, avoiding repeated distribution-function projections between different kinetic representations. As a proof of concept, a multiscale Couette-flow simulation is shown in Fig. \ref{fig: Multiscale}. The central region uses a $D2Q3^2$ velocity set constructed from full-range Gauss-Hermite quadrature nodes, while the near-wall regions use a $D2Q4^2$ velocity set constructed from half-range Gauss-Hermite quadrature nodes. When information is transferred from low-order to high-order moment representation, the missing high-order equilibrium moments are exact Maxwell moments, and nonequilibrium moments are reconstructed using Eq. \eqref{eq19}; when information is transferred from high-order to low-order representation, the required primary moments are obtained directly. This successful coupling illustrates the potential of the present moment framework for locally adaptive velocity sets. 

\textit{Acknowledgments}-
Part of this research was supported by the Alexander von Humboldt Foundation, Germany; and by the French project COLIBRI funded by DGAC and supported by Next generation EU in the frame of "Plan national de Relance et de Résilience français (PNRR)"; and by the ANR, Airbus, Fives-Pillard and SafranTech through the Industrial Chair Program LIBERTY ANR-23-CHIN-0005. The first author also thanks Fengye Wei for helpful discussions.

\appendix

\nocite{*}
\bibliographystyle{apsrev4-2}
\bibliography{Ref3}

\end{document}